\documentclass[preprint,aps]{revtex4}%
\usepackage{amsfonts}
\usepackage{amsmath}
\usepackage{amssymb}
\usepackage{graphicx}%
\providecommand{\U}[1]{\protect\rule{.1in}{.1in}}

\begin{document}
\title{Influence of electron scatterings on thermoelectric effect}
\author{Jing Li}
\email{jing.li.phy@gmail.com}
\author{Tin Cheung Au Yeung}
\author{Chan Hin Kam}
\affiliation{Microelectronic Division, School of Electric and Electronic Engineering,
Nanyang Technological University, Singapore 639798, Singapore}
\pacs{PACS number}

\begin{abstract}
In this work, we employed non-equilibrium Green's function to investigate the
electron transport properties in the nanowire with the presence of
scatterings. The scattering mechanism is modelled by using the concept of
B\"{u}ttiker probe. The effect of electron scattering is analyzed under three
conditions: absence of external field; with a bias voltage; and with a finite
temperature difference. It is found weak and strong scatterings strength
affect the electron transport in different ways. In the case of weak
scattering strength, electron trapping increase the electron density,
hereafter boost the conductance significantly. Although the increment in
conductance would reduce the Seebeck coefficient slightly, the power factor
still increases. In the case of strong scattering strength, electron
diffraction causes the redistribution of electrons, accumulation of electron
at the ends of the wire blocks current flow; hence the conductance is reduced
significantly. Although the Seebeck coefficient increases slightly, the power
factor still decreases. The power factor is enhanced by $6\%-18\%$, at the
optimum scattering strength.

\end{abstract}
\volumeyear{year}
\volumenumber{number}
\issuenumber{number}
\eid{identifier}
\date[Date text]{date}
\received[Received text]{date}

\revised[Revised text]{date}

\accepted[Accepted text]{date}

\published[Published text]{date}

\startpage{1}
\endpage{2}
\maketitle

\section{Introduction}

Thermoelectric (TE) device is a type of green energy devices and is of great
possibility to be used widely in near future. Currently, the limitation of TE
device is its low energy conversion efficiency, and has attracted much
attention \cite{Science1999, Nature2001, Science2002, Science2004,
Nature2008YangPD, 159Nature2008, Ali}.  The efficiency of a TE device is
indicated by the dimensionless TE figure of merit ($ZT$), determined by the
expression $S^{2}\sigma T/\kappa$ using its parameters Seebeck coefficient
$S$, electric conductivity $\sigma$, thermal conductivity $\kappa$
($=\kappa_{e}+\kappa_{ph}$, contributed by electrons and phonons) and the
temperature $T$. Among these parameters, $S$, $\sigma$ and $\kappa_{e}$ are
related to the electron transport and $\kappa_{ph}$ is related to the phonon
transport. Therefore, the strategies of developing high $ZT$ device are
improving the electronic TE efficiency and reducing the lattice thermal
conductivity\cite{LJTE}. Currently, most of the efforts are put in to reduce
the lattice thermal conductivity by investigation the phonon transport, which
can be found in our previous works \cite{LJ1, LJ2, Zhao}. However, more
attentions needs to be paid on the electron transport, as the electronic TE
efficiency limits the performance of TE device significantly\cite{LJTE}. The
best TE device has a Dirac delta shaped electron transmission
function\cite{Mahan 1996}. Besides the TE efficiency, the power factor
($S^{2}\sigma$), related to the power rating, is another important index of TE device.

In nano-scaled device, the concept of ballistic transport of electrons is
important. However, electron scatterings still occurs for the reason of
defects, grain boundaries, phonons, and charges in the device. Electron
scatterings affect the transport properties, and further influence the TE
effect of the device. A previous work has studied the effect of nano-particle
scattering on TE power factor\cite{Mona}. It is found that the electron
concentration is usually higher in the sample with nano-particles, implying
the Seebeck coefficient is usually unchanged and conductivity is increased at
the optimum of the power factor. That work is based on low nano-particle
concentration, and that work rise our interest in the problem of the electron
scattering strength on TE effect. In this work, we employed non-equilibrium
Green's function to investigate the electron transport properties in nanowire
with the presence of electron scatterings. Scattering points are inserted into
the nanowire, and are modelled by B\"{u}ttiker probes\cite{Datta2003}. The
response of the device to an external field, due to a temperature difference
or a bias voltage, is discussed with a wide range of scattering strength.

\section{Theory}

A silicon nanowire with square cross-sectional shape coupled by two contacts,
which are served as thermal and electric reservoirs, is considered as shown in
Fig.1. The wire is composed of many slices with the thickness ($a_{z}$)
$0.3$nm parallel to the contact interface. For the slice at position $z$, the
effective potential is given by%
\begin{equation}
E_{n_{1},n_{2}}(z)=\frac{\hbar^{2}\pi^{2}}{2}[\frac{1}{m_{x}^{\ast}}%
(\frac{n_{1}}{l_{x}})^{2}+\frac{1}{m_{y}^{\ast}}(\frac{n_{2}}{l_{y}}%
)^{2}]+V(z)\text{.}%
\end{equation}
The variables $n_{1,(2)}$ is the quantum number for the energy in transverse
direction; $l_{x,(y)}$ the side length of the cross-section (set to $5$nm in
the computation); $V(z)$ the electric potential energy; $m^{\ast}$ the
effective mass of silicon; and $\hbar$ the Dirac constant. The Hamiltonian of
the wire for the mode, denoted by the subscript ($n_{1},n_{2}$), is
constructed as follow:
\begin{equation}
H=\left[
\begin{array}
[c]{cccc}%
_{E_{n_{1},n_{2}}(1)+2t_{z}} & _{-t_{z}} & _{0} & \ldots\\
_{-t_{z}} & _{E_{n_{1},n_{2}}(2)+2t_{z}} & _{-t_{z}} & \ldots\\
_{0} & -t_{z} & \ddots & \ldots\\
\vdots & \vdots & \vdots & \ddots
\end{array}
\right]  \text{,}\label{Hamiltonian}%
\end{equation}
where $t_{z}$ is the coupling energy between two slices determined by the
expression $\hbar^{2}/(2m_{z}^{\ast}a_{z}^{2})$ (about $461$meV in this model).%

\begin{figure}
[ptb]
\begin{center}
\includegraphics[
height=1.4589in,
width=3.1981in
]%
{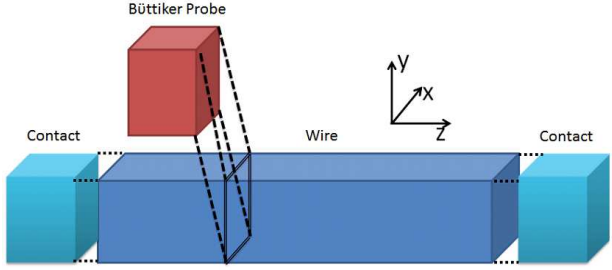}%
\caption{The schematic of the modelled system: a nanowire coupled by two
contacts at both ends, a scattering point denoted by the square box is coupled
to B\"{u}ttiker probe through a virtual contact.}%
\end{center}
\end{figure}

To reveal the effect of the electron scattering, the scattering points
modelled by B\"{u}ttiker probes are inserted into the wire. At a scattering
point, the slice of the wire is coupled to the B\"{u}ttiker probe with the
coupling strength ($U$),\ as shown in Fig.1. The coupling strength determines
the electron scatterings strength. For the convenience in discussion, we
define the scattering factor $\alpha$, the relative scattering strength
respected to the coupling strength between two slices. Hence $U=\alpha\cdot
t_{z}$, and $\alpha$ equal to $0.056$ is equivalent to the thermal energy at
$300$K, which is about $26$meV.

All the contacts in the system, including both virtual contacts (B\"{u}ttiker
probes) and two real contacts, are labeled by index $i$. The self-energy due
to the contact $i$ is given by%
\begin{equation}
\Sigma_{i}(E)=U^{2}\cdot\frac{-e^{i\cdot k_{i}(E)\cdot a_{z}}}{t_{z}\cdot
a_{z}}\text{,}\label{Self-Energy}%
\end{equation}
where $k_{i}(E)$ is the longitudinal wave vector determined by
\begin{equation}
k_{i}(E)=\sqrt{\frac{2m_{z}^{\ast}}{\hbar^{2}}[E-E_{n_{1},n_{2}}(z)]}%
\text{.}\label{k(E,z)}%
\end{equation}
The broadening function of the contact $i$ is then obtained from its
self-energy:
\begin{equation}
\Gamma_{i}(E)=-2\operatorname{Im}[\Sigma_{i}(E)]\text{.}%
\end{equation}
The total self-energy of the wire $\Sigma$ is the sum of the contribution from
each contact. The retarded and advanced Green's function of the wire, $G^{R}$
and $G^{A}$, are then determined by:
\begin{align}
G^{R}(E) &  =[E+i\epsilon-H-\Sigma(E)]^{-1}\text{,}\\
G^{A}(E) &  =[G^{R}(E)]^{\dag}\text{,}%
\end{align}
where $\epsilon$ is an infinitesimal positive number. With that, the electron
transmission function between contact $i$ and contact $j$ is given by%
\begin{equation}
T_{i,j}(E)=Tr[\Gamma_{i}G^{R}\Gamma_{j}G^{A}]\text{.}\label{Transmision}%
\end{equation}
The electric current of contact $i$ is determined by the Landauer-B\"{u}ttiker
formula:%
\begin{equation}
I_{i}=\frac{2e}{h}%
{\displaystyle\int}
dE%
{\displaystyle\sum\limits_{j\neq i}}
T_{i,j}\cdot(f_{i}-f_{j})\text{,}\label{Current}%
\end{equation}
where $f_{i}$ is the Fermi-Dirac distribution at contact $i$ with the Fermi
level $E_{f,i}$, defined by
\begin{equation}
f_{i}=[e^{\frac{E-E_{f,i}}{k_{B}\cdot T}}+1]^{-1}\text{.}%
\end{equation}
To fulfill the charge conservation, the net current of each B\"{u}ttiker probe
is set to zero by adjusting the Fermi levels of B\"{u}ttiker probes using
Jacobian method, (linear temperature gradient is assumed in the wire). The
Fermi levels in the left and right contacts are determined by assuming all
donors are ionized, and the donor concentration $10^{18}$cm$^{-3}$ is used in
the computation. The number of electrons in each slices of the wire are
obtained from the diagonal elements of $n$, determined by the following
expression:%
\begin{equation}
n=\frac{1}{\pi}\int dE\sum_{i}f(E,E_{f,i})G^{R}\Gamma_{i}G^{A}\text{.}%
\end{equation}
The potential energy $V(z)$ for each slice is updated by the Poisson equation
using the electron density determined by the above formula, and the final
status of the system is obtained after the self-consistent iteration process.

\section{Discussion}

In the following, the system at room temperature in three conditions: (A)
absence of external field, (B) with a bias voltage, and (C) with a finite
temperature difference, are discussed.

\subsection{Absence of external field: electron trapping and diffraction}

In this section, the length of the nanowire is set to $18$nm, and two
scattering points are inserted at the positions of $z=6$nm and $z=12$nm. With
the absence of external field, i.e. both temperatures and voltages at the two
contacts are equal, the electron charge density along the wire determined by
Green's function is shown in Fig.2, in which the scattering points are denoted
by \textquotedblleft SP1\textquotedblright\ and \textquotedblleft
SP2\textquotedblright. In the case of ballistic transport ($\alpha=0$),
electron scattering only occurs at the contact interfaces, but not in the
wire. For this reason, electrons are accumulated at both ends of the wire,
that incur the repulsion of electron as the result of the Coulomb force.
Therefore, a fluctuation of electron density is observed in the wire, as shown
by the \textquotedblleft Ballistic\textquotedblright\ curve in Fig.2a. When
the scattering strength getting slightly stronger ($\alpha<0.1$), the electron
wave function in the wire is reconstructed due to the perturbation of the
scattering points, but this effect is not significant. Electron scattering
occurs when electron collide with the scattering points, and electrons are
accumulated in the wire. Fig.2a shows the stronger scattering strength, the
more electrons are trapped inside the wire, especially near the scattering
points in the weak scattering strength region ($\alpha<0.1$). This result is
consistent with the previous work\cite{Mona}. However, in strong scattering
strength region, this effect vanished. This is because the electron wave
function is significantly modified by the scattering points. The probability
of finding an electron near the scattering points become much smaller,
therefore the electron density near scattering points reduces (Fig.2b), the
chance of electron scatterings decreases, the chance of electron scatterings
decreases. This phenomenon is known as the electron diffraction, electrons
prefer avoiding the scattering points than being scattered. %

\begin{figure}
[ptb]
\begin{center}
\includegraphics[
height=4.3223in,
width=3.1981in
]%
{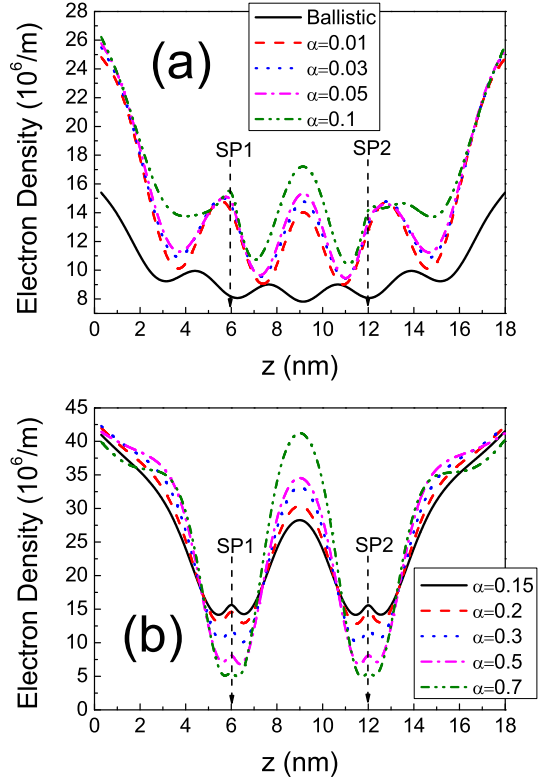}%
\caption{The electron density in a nanowire with the length $18$nm, two
scattering points are inserted at $z=6$nm and $z=12$nm, (a) is for the weak
scattering strength region and (b) is for the strong scattering strength
region. $\alpha$ is the relative scattering strength.}%
\end{center}
\end{figure}

\subsection{\bigskip A bias voltage}

In the following, we discussed a $9$nm nanowire at room temperature with a
bias voltage. A single scattering point is considered at the center of the
wire. In the case of ballistic transport ($\alpha=0$), due to the external
field, electron density is tilting toward one end of the wire, as shown in
Fig.3a. With slightly increases in the scattering strength, more electrons are
accumulated in the wire, especially near the scattering point. This is
consistent with the discussion in section A. With the bias voltage, the
increment in electron density is conducive to boost the electric current as
shown by Fig.3b. However, in strong scattering strength region, electrons are
bypassing the scattering point by reconstructing their wave functions. This
leads to the accumulation of electrons near the contact interfaces (Fig.3c),
which block current flow. So, decrease in electric current is expected (see
Fig.3d).%
\begin{figure}
[ptb]
\begin{center}
\includegraphics[
height=2.341in,
width=3.1981in
]%
{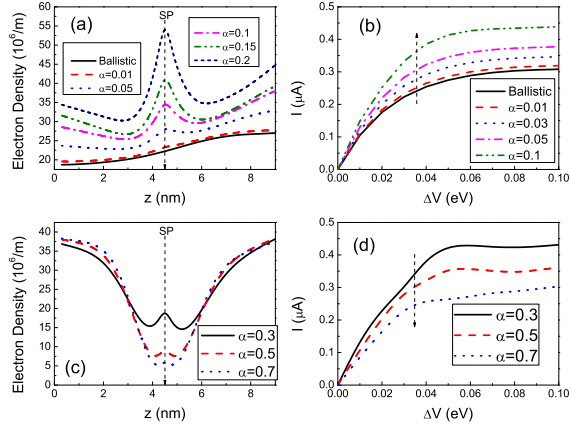}%
\caption{A $9$nm nanowire with the presence of bias voltage $\Delta V$, (a):
the electron density, and (b): the electric current in weak scattering
strength region; (c): the electron density, and (d): the electric current in
strong scattering strength region. $\alpha$ is the relative scattering
strength.}%
\end{center}
\end{figure}

With different bias voltages, the electric current respect to the scattering
strength are in the similar profile: a rise followed by a drop (Fig.4). Also,
the scattering strength at which the maximum current occurs, increases with
the bias voltage. This phenomenon can be understood in the following way. The
electrons are trapped in the wire in the weak scattering strength region. With
the increase in the bias voltage, these electrons can be accelerated and
contribute to the electric current. Higher bise voltage has stronger power in
accelerating electron, therefore the scattering strength at which the maximum
current occurs is increasing with the bias voltage. With sufficiently large
scattering strength, the diffraction of electron causes the accumulation of
electrons at both ends of the wire, which blocks the electric current.%

\begin{figure}
[ptb]
\begin{center}
\includegraphics[
height=2.2554in,
width=3.1981in
]%
{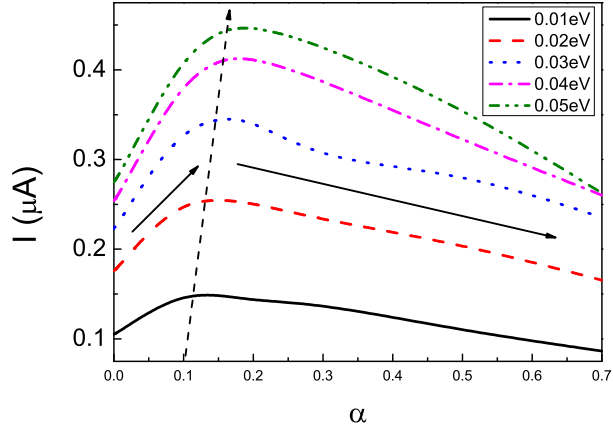}%
\caption{A $9$nm nanowire with various bias voltages, the electric current is
plotted with respect to the relative scattering strength $\alpha$.}%
\end{center}
\end{figure}

\subsection{A finite temperature difference}

To discussion the TE effect, a $9$nm nanowire under a temperature difference
is considered. The temperature difference is split into half and applied to
the contacts, i.e. $T_{L}=T+\Delta T/2$ and $T_{R}=T-\Delta T/2$. In general,
TE effect is evaluated in the open circuit condition. Therefore, the net
current is zero. For this reason, to offset the current produced by the
temperature difference, an electric field is generated. The ratio of the
voltage difference over the temperature difference gives the Seebeck
coefficient. To determine the conductance of the wire in the presence of
temperature difference, a testing current (a small current) is injected into
the wire. From the change of the voltage difference, the conductance is
obtained.%
\begin{figure}
[ptb]
\begin{center}
\includegraphics[
height=2.3687in,
width=3.1981in
]%
{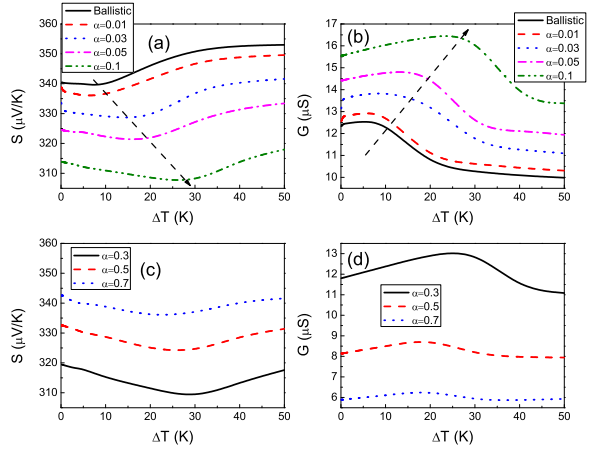}%
\caption{A $9$nm nanowire with temperature difference $\Delta T$, (a): Seebeck
coefficient, and (b): electric conductance in weak scattering strength region;
(c): Seebeck coefficient, and (d): electric conductance in strong scattering
strength region. $\alpha$ is the relative scattering strength.}%
\end{center}
\end{figure}

In the weak scattering strength region ($\alpha<0.1$), Seebeck coefficient is
smaller with stronger scattering. That is because the electric conductance
increases (Fig.5b), and a smaller electric field is sufficient enough to
offset the current produced by temperature difference for case of larger
conductance. Similarly, in the strong scattering strength region, Seebeck
coefficient increases as the result of the decrease in the electric
conductance, as shown in Fig.5c and Fig.5d.

In Fig.5b, the conductance increases with the temperature difference, and
starts to decrease after the critical temperature difference, also the
critical temperature difference increases with the scattering strength. This
is because, when temperature difference is small, the increment in temperature
difference intends to inject more electrons into the wire; these electrons are
trapped easily by the scattering point, lead to the a higher electron density
and a larger conductance. When the temperature difference is more than the
critical value, more electrons are injected than the wire could conduct, that
causes the redistribution of electrons, which limits current flow. Due to this
effect on conductance, Seebeck coefficient decreases at first and start to
increase after the critical temperature difference (Fig.5a). In the strong
scattering strength region, the fluctuations in conductance and Seebeck
coefficient are not as much as that in weak scattering strength region. This
is because the chance of electron scatterings is reduced due to the effect of
electron diffraction, number of electrons trapped in the wire is smaller than
the case of weak scattering strength.
\begin{figure}
[ptb]
\begin{center}
\includegraphics[
height=1.2194in,
width=3.1981in
]%
{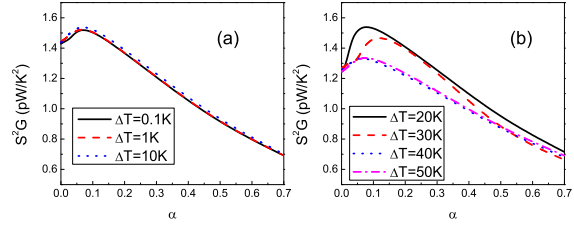}%
\caption{A $9$nm nanowire with temperature difference $\Delta T$, the
thermoelectric power factor is plotted with respect to the relative scattering
strength $\alpha$ with (a): small temperature difference, and (b): large
temperature difference.}%
\end{center}
\end{figure}

Fig.6 shows the TE power factors with the change of scattering strength. The
power factor has maximum value at about $\alpha=0.1$, and it drops
dramatically after that. The explanation of the change in power factor can be
ascribed to the change in the the conductance and Seebeck coefficient.  At the
optimum power factor ($\alpha=0.1$), comparing to the ballistic case, for
$\Delta T$ equals to $0$K, $20$K and $50$K,  the Seebeck coefficient drops by
$8\%$, $13\%$ and $11\%$ (Fig.5a); but the conductance increases by $25\%$,
$57\%$ and $35\%$ (Fig.5b), therefore the power factor is increased by $6\%$,
$18\%$ and $7\%$, respectively. Due to the shifting of the critical
temperature difference mentioned in previous paragraph, the power factor has
the maximum enhancement when the temperature difference is about $20$K.

\section{Conclusion}

Weak and strong scatterings strength affect the electron transport in
different manners. In the case of weak scattering strength, electron trapping
increase the electron density, boost the conductance significantly. The
increment in conductance would reduce the Seebeck coefficient slightly, but
overall the power factor increases. In the case of strong scattering strength,
electron diffraction causes the redistribution of electrons, accumulation of
electrons at the ends of the wire block current flow, hence the conductance is
suppressed significantly. Due to this effect, the Seebeck coefficient
increases slightly, but the power factor decreases severely.


\begin{thebibliography}{99}                                                                                               %


\bibitem {Science1999}Francis J. DiSalvo, Science \textbf{285}, 703 (1999).

\bibitem {Nature2001}Rama Venkatasubramanian, Edward Siivola, Thomas Colpitts,
and Brooks O'Quinn, Nature \textbf{413}, 597 (2001).

\bibitem {Science2002}T. C. Harman, P. J. Taylor, M. P. Walsh, and B. E.
LaForge, Science \textbf{297}, 2229 (2002).

\bibitem {Science2004}Kuei Fang Hsu, Sim Loo, Fu Guo, Wei Chen, Jeffrey S.
Dyck, Ctirad Uher, Tim Hogan, E. K. Polychroniadis, and Mercouri G.
Kanatzidis, Science \textbf{303}, 818 (2004).

\bibitem {Nature2008YangPD}Allon I. Hochbaum, Renkun Chen, Raul Diaz Delgado,
Wenjie Liang, Erik C. Garnett, Mark Najarian, Arun Majumdar, and Peidong Yang,
Nature \textbf{451}, 163 (2008).

\bibitem {159Nature2008}Akram I. Boukai, Yuri Bunimovich, Jamil Tahir-Kheli,
Jen-Kan Yu, William A. Goddard Iii, and James R. Heath, Nature \textbf{451},
168 (2008).

\bibitem {Ali}Sebastian Volz, Ali Shakouri, and Mona Zebarjadi, in Thermal
Nanosystems and Nanomaterials (Springer Berlin / Heidelberg, 2009), Vol.
\textbf{118}, pp. 225.

\bibitem {Mahan 1996}G. D. Mahan and J. O. Sofo, Proceedings of the National
Academy of Sciences \textbf{93}, 7436 (1996).

\bibitem {Mona}Mona Zebarjadi, Keivan Esfarjani, Ali Shakouri, Je-Hyeong Bahk,
Zhixi Bian, Gehong Zeng, John Bowers, Hong Lu, Joshua Zide, and Art Gossard,
Applied Physics Letters \textbf{94}, 202105 (2009).

\bibitem {Datta2003}R. Venugopal, M. Paulsson, S. Goasguen, S. Datta, and M.
S. Lundstrom, Journal of applied physics \textbf{93}, 5613 (2003).

\bibitem {LJ1}Jing Li, T. C. Au Yeung, C. H. Kam et al, J. Appl. Phys,
\textbf{106, }054312 (2009).

\bibitem {LJ2}Jing Li, T. C. Au Yeung, C. H. Kam et al, J. Appl. Phys,
\textbf{106, }014308 (2009).

\bibitem {Zhao}X. Zhao et al., Journal of Applied Physics \textbf{107}, 094312 (2010).

\bibitem {LJTE}Jing Li, T. C. Au Yeung, and C. H. Kam, J. Phys. D: Appl. Phys.
\textbf{45}, 085102 (2012).
\end{thebibliography}
\end{document}